\documentclass[a4paper]{article}
\usepackage[utf8]{inputenc}
\usepackage[margin=3cm]{geometry}
\usepackage{setspace}
\usepackage{lineno}

\usepackage{graphicx}
\usepackage{xcolor}
\usepackage{url}

\title{Comment on Buss \emph{et al.}, \emph{Science} 2021:\\ An alternative, empirically-supported adjustment for sero-reversion yields a 10 percentage point lower estimate of the cumulative incidence of SARS-CoV-2 in Manaus by October 2020}
\author{Sarah Kadelka,$^{1\ast}$ Judith A. Bouman$^{1}$, Peter Ashcroft$^{1}$, Roland R. Regoes$^{1\ast}$\\
\normalsize{$^{1}$Department of Environmental Systems Science, Institute of Integrative Biology,}\\
\normalsize{ETH Z\"urich, Z\"urich, Switzerland }\\ 
\normalsize{$^\ast$E-mail:  sarah.kadelka@env.ethz.ch, roland.regoes@env.ethz.ch}}
\date{\today}

\begin{document}

\maketitle

\begin{abstract}
The estimate of the cumulative incidence of SARS-CoV-2 of 76\% in Manaus by October 2020 by Buss \emph{et al.} relies on the assumption of an exponentially-declining probability of sero-reversion over time.
We present an alternative, empirically-supported approach that is based on the observed dynamics of antibody titers in sero-positive cases.
Through this approach we revise the cumulative incidence estimate to 66\% (63.3\% - 68.5\%) by October 2020.
This estimate has implications for the proximity to herd immunity and future estimates of fitness advantages of virus variants, such as the P.1 variant of concern.
This methodology is also relevant for any other sero-survey analysis that has to adjust for sero-reversion.
\end{abstract}

\onehalfspacing
\setlength{\parskip}{0.5em}

In their paper on the attack rate of SARS-CoV-2 in the Brazilian Amazon, Buss \emph{et al.} estimated a cumulative incidence of 76\% (95\% CI, 66.6\% to 97.9\%) in Manaus by October 2020 \cite{buss2021three}.
While this level of past infection should have conferred herd immunity and curbed the epidemic \cite{anderson2020will,fontanet2020covid,britton2020mathematical}, new infections with SARS-CoV-2 are still reported and COVID-19 related deaths are accumulating to-date 
\cite{sabino2021resurgence}.
Multiple explanations for these puzzling developments have been proposed: methodological issues relating to cumulative incidence estimation, waning of immunity, and new viral variants that evade immunity from previous infection or have increased transmissibility compared to the initially circulating variant \cite{sabino2021resurgence}.

Here we report on methodological issues which suggest that the cumulative incidence estimate of 76\% in Manaus represents an over-estimation.
Buss \emph{et al.} estimated the cumulative incidence by adjusting the raw sero-prevalence estimate for the sensitivity and specificity of the serological test and for sero-reversion. 
Sero-reversion occurs because SARS-CoV-2-specific antibody titers, after peaking approximately 3--4 weeks after infection, decline exponentially over time \cite{roltgen2020defining,dan2021immunological,antia2018heterogeneity}.
Once the antibody titer decreases below the sero-positivity threshold, the study participant `reverts' from sero-positive to sero-negative.

\begin{figure}[h!]
\centering
\includegraphics[width = 1\textwidth]{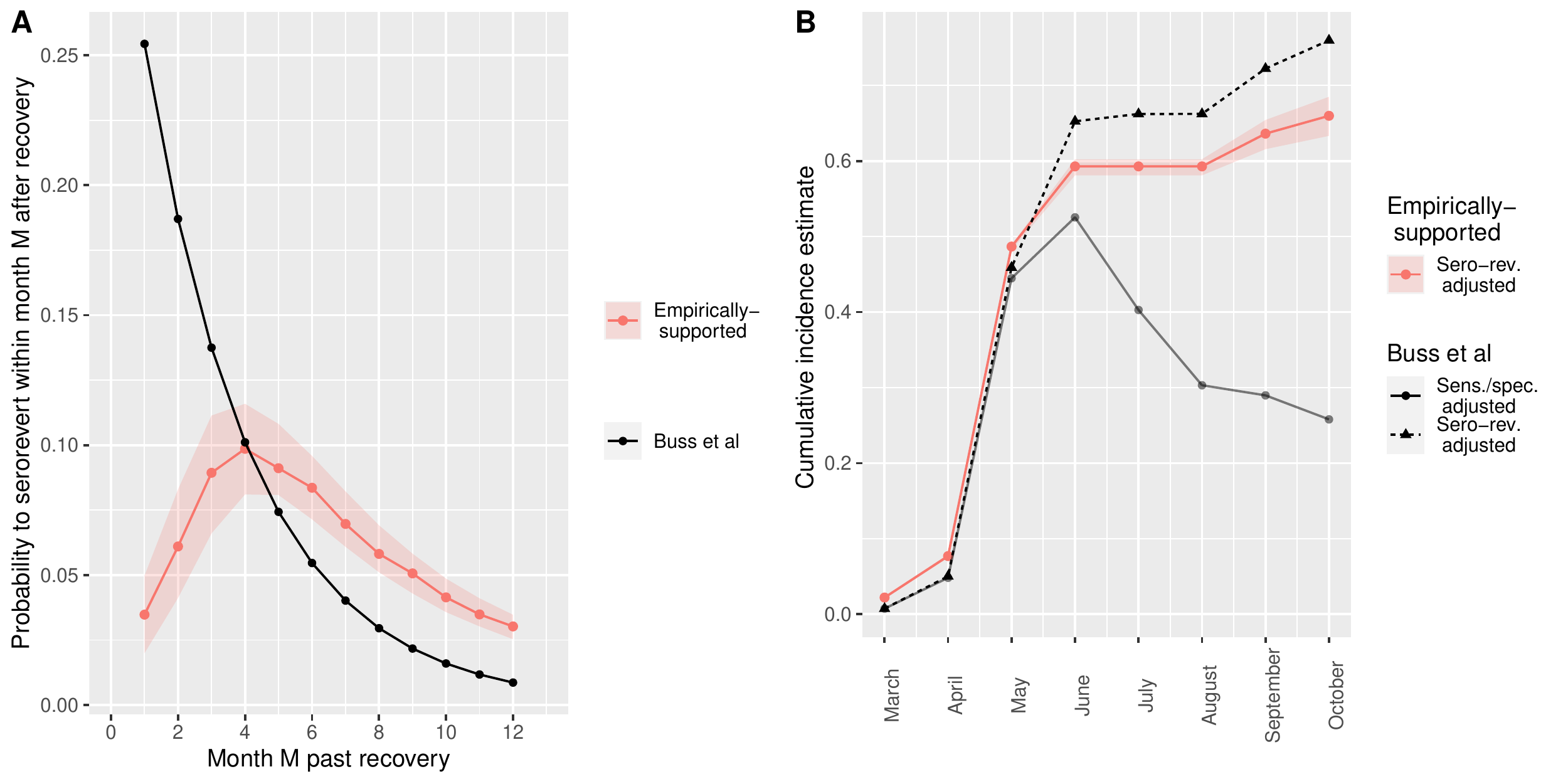}
\caption{\textbf{(A)} Probabilities to sero-revert within a given month after recovery assuming an exponential decay of these probabilities as proposed by Buss \emph{et al.} (black). Probabilities to sero-revert within a given month after recovery (=antibody peak) assuming an exponential decay of antibodies (red). 
\textbf{(B)} Cumulative incidence estimates adjusted for test accuracy (black dots) and additionally for sero-reversion using Buss \emph{et al.}'s probabilities of sero-reversion (black triangles). Sensitivity, specificity and sero-reversion adjusted cumulative incidence estimates using the empirically-derived sero-reversion probabilities assuming exponential decay of antibodies (red). The shaded areas show the 95\% confidence intervals for the respective point estimates.}
\label{fig:Figure1_combined}
\end{figure}

To adjust for sero-reversion, Buss \emph{et al.} assumed that the probability of sero-reversion declines exponentially over time (see Figure~\ref{fig:Figure1_combined}A, black curve).
While seemingly consistent with the exponential decline of antibody titers, we show that this assumption over-corrects for sero-reversion, in particular shortly after the antibody peak which we assume to coincide with recovery.
Using the data on antibody titers and their temporal declines provided by Buss \emph{et al.}, we derived an empirically-supported profile for the probability of sero-reversion. 
In brief, we considered the observed distribution of antibody titers from 193 positive controls and simulated their exponential decline based on antibody decay rates sampled from 81 convalescent plasma donors.
From this ensemble of antibody titer trajectories, we determined the probability to sero-revert as a function of time since antibody peak \cite{kadelka2021gitlab}.

In contrast to the exponential profile used by Buss \emph{et al.}, we find that the sero-reversion probability first increases before dropping after four months past antibody peak, 
(see Figure~\ref{fig:Figure1_combined}A, red curve).
When correcting the cumulative incidence estimate for sero-reversion using the empirically-supported profile, we obtain a cumulative incidence estimate of 66\% (bootstrap 95\% CI, 63.3 to 68.5\%) in October 2020
--- 10 percentage points lower than the original estimate of 76.0\% and outside the 95\% confidence interval (66.6\% to 97.9\%), i.e. significantly lower (see Figure~\ref{fig:Figure1_combined}B, red dots).

Using the available antibody data in the paper, we have shown that Buss et al. introduce a significant over-estimation of the cumulative incidence of SARS-CoV-2 in Manaus. Currently, the positive control data we used to derive the sero-reversion probability is only representative of symptomatic, non-hospitalised COVID-19 cases. However, antibody levels vary with disease severity \cite{hashem2020early}, therefore time to sero-reversion might vary as well. Additional positive control data of mild and asymptomatic cases could improve our empirical model even further and potentially change the estimated sero-reversion-corrected cumulative incidence.

Similar to Buss \emph{et al.}, we estimated the cumulative incidence under the assumption that incidence can only increase.
This assumption, while consistent with the concept of cumulative incidence, ignores potential uncertainties in the seroprevalence estimates and could lead to a further over-estimation of the cumulative incidence.
A case in point is the observed drop in the sensitivity and specificity adjusted prevalence from 52.5\% in June to 40.3\% in July.
This observed decrease is faster than can be explained by exponential antibody decay at rates reported by Buss \emph{et al.}, and would require antibody half-lives shorter than 40 days --- contradicting previously reported IgG half-lives of 50--106 days \cite{dan2021immunological}.
Thus, there are remaining inconsistencies in these data that need to be resolved in the future.

Even though the alternative, empirically-supported sero-reversion correction does not provide a full explanation for the unexpected resurgence of the SARS-CoV-2 epidemic in Manaus, it contributes to solving the puzzle by providing a lower estimate of the cumulative incidence.
This lower estimate of the cumulative incidence in Manaus will also impact recent estimates of the transmissibility and the extent of immune evasion of the novel variant of concern, lineage P.1 \cite{faria2021genomics}.
Beyond its relevance for the Brazilian sero-survey, the approach to adjusting for sero-reversion presented here provides an important, empirically-supported method that could be used in any sero-survey in which the antibody levels wane over time.

We would like to emphasize that this contradicting observation was only possible because the authors of the original study generously and comprehensively shared their data.

\bibliographystyle{plain}
\bibliography{references}
\end{document}